\documentclass[conference]{IEEEtran}
\makeatletter
\def\ps@headings{%
\def\@oddhead{\mbox{}\scriptsize\rightmark \hfil \thepage}%
\def\@evenhead{\scriptsize\thepage \hfil \leftmark\mbox{}}%
\def\@oddfoot{}%
\def\@evenfoot{}}
\makeatother 
\usepackage{cite}

\ifCLASSINFOpdf
   \usepackage[pdftex]{graphicx}
\else
   \usepackage[dvips]{graphicx}
\fi
\usepackage{algorithmic}
\usepackage{algorithm}
\usepackage[tight,footnotesize]{subfigure}
\begin{document}
%
\title{Approximation Algorithms for Link Scheduling with Physical Interference Model in Wireless Multi-hop Networks}



\author{\IEEEauthorblockN{Shuai Fan}
\IEEEauthorblockA{Department of Electronic Engineering\\
Tsinghua University\\
Beijing, P.R. China\\
fans05@mails.tsinghua.edu.cn} \and \IEEEauthorblockN{Lin Zhang}
\IEEEauthorblockA{Department of Electronic Engineering\\
Tsinghua University\\
Beijing, P.R. China\\
Email: linzhang@tsinghua.edu.cn} \and \IEEEauthorblockN{Yong Ren}
\IEEEauthorblockA{Department of Electronic Engineering\\
Tsinghua University\\
Beijing, P.R. China\\
Email: reny@tsinghua.edu.cn}}


%


\maketitle

\begin{abstract}
The link scheduling in wireless multi-hop networks is addressed.
Different from most of work that adopt the protocol interference
model which merely take consideration of packet collisions, our
proposed algorithms use the physical interference model to reflect
the aggregated signal to interference and noise ratio (SINR), which
is a more accurate abstraction of the real scenario. We first
propose a centralized scheduling method based on the Integer Linear
Programming (ILP) and resolve it by an approximate solution based on
the randomized rounding method. The probability bound of getting a
guaranteed approximate factor is given. We then extend the
centralized algorithm to a distributed solution, which is favorable
in wireless networks. It is proven that with the distributed
scheduling method, all links can transmit without interference, and
the approximate ratio of the algorithm is also given.
\end{abstract}


%
\IEEEpeerreviewmaketitle

\section{Introduction}\label{section1}
%
%
In wireless multi-hop networks, concurrent transmissions that share
a common channel may cause interference, and if too many devices
transmit simultaneously, the interference will prevent an intended
receiver from receiving the signal, and causes message loss. On the
other hand, if too few nodes transmit at the same time, valuable
bandwidth is wasted and the overall throughput may degenerate.
Hence, the classic problem faced by the MAC layer scheduling
protocols is to select an appropriate set of devices for concurrent
transmissions, so that the interference does not cause message loss.
In slotted wireless multi-hop networks, a natural and important goal
of scheduling algorithms is to maximize the total throughput with
the interference restriction.

Hence, an accurate model of interference is fundamental in order to
derive theoretical or simulation-based results. There are two main
interference models used, namely the protocol interference model and
the physical interference models \cite{gupta2000cwn}. In the
protocol model, a communication from node $u$ to node $v$ is
successful if no other node within a certain interference range from
$v$ is simultaneously transmitting. Due to its simplicity and to the
fact that it can be used to mimic the behavior of CSMA/CA networks
such as IEEE 802.11, this model has been widely used in the
literature. However, it doesn't reflect the advanced physical layer
technologies that allow concurrent multiple signal reception. In the
physical interference model, a communication between nodes $u$ and
$v$ is successful if the Signal to Interference and Noise Ratio
(SINR) at receiver $v$ is above a certain threshold. This model is
less restrictive than the protocol interference model, and thus
higher network capacity can be achieved by applying the physical
interference model.

Clearly, the interference in the protocol model is a tremendous
simplification of the physical reality faced in the wireless
multi-hop networks. Particularly, the interference caused by
different transmitters may accumulate and is not binary, i.e., does
not vanish beyond any specific border. Moreover, a node may
successfully receive a message even when there are other
concurrently transmitting nodes in its interference range. In
\cite{behzad2003pgb,gronkvist2001cbg}, it is argued that the
performance of protocol model based algorithms is inferior to those
with more realistic and fundamental physical interference model.
More recently, Moscibroda et al. \cite{moscibroda2006pdb} show
experimentally that the theoretical limits of any protocol, which
obeys the laws of protocol interference model, can be broken by a
protocol explicitly defined for the physical interference model.
However, although the physical interference model enjoy a much
better performance, the scheduling problem under this model is
notoriously hard to resolve.

In this paper, we study the link scheduling problem under the
physical interference model. First, we formulate the problem as an
Integer Linear Programming (ILP) problem, which is similar to the
work in \cite{friderikos2006non}. Since it is NP-complete
\cite{goussevskaia2007complexity}, we get the approximate algorithm
by using the randomized rounding method, and the probability bound
of getting a guaranteed approximate ratio is given. Unlike the work
in \cite{friderikos2006non}, we adjust the results of the rounding
procedure based on all the constraints, and no iteration procedure
is needed.

Since the centralized algorithm needs the global information of the
network to do the scheduling, including the location of all nodes,
the load of the links to be scheduled in a certain time slot, etc.,
collecting the required information may causes a lot of bandwidth
overhead, especially when the network scale is large. Moreover, in
some cases, the information is hard to get. The difficulties in the
implementation of centralized algorithm call for the distributed
scheduling scheme that may achieve a fraction of the overall optimal
performance. We then propose a distributed approach based on
physical interference model. Each sending node implements the
approach in three phases at each time slot, and all the constraints
of the ILP problem in the centralized algorithm are satisfied. To
analyze the performance of the algorithm, the approximation ratio is
given.

The rest of the paper is organized as follows. After giving an
overview of related work in Section \ref{section2}, the network and
interference model is described in Section \ref{section3}. In
Section \ref{section4}, scheduling problem under physical
interference model is formulated and we present the centralize
algorithm to get the approximate solution. In section
\ref{section5}, a distributed approach is proposed. Section
\ref{section6} presents the simulation results of the proposed
algorithmss and Section \ref{section7} concludes the paper.

\section{Related work}\label{section2}
The problem of scheduling link transmissions in a wireless network
in order to optimize one or more of performance objectives (e.g.
throughput, delay, fairness or energy) under interference
constraints has been a subject of much interest over the past
decades.

As mentioned in the previous section, an accurate modeling of
interference is fundamental to the scheduling problem. Most of the
scheduling mechanisms proposed for wireless multi-hop networks use
the protocol interference model, such as
\cite{hajek1988lsp,kumar2004eep,moscibroda2005cur,ramanathan1993sam,sharma2006csw}.
These algorithms usually employ an implicit or explicit coloring
strategy, and simply neglect the aggregated interference of nodes
located farther away. However, the interference caused by a
transmitter is not binary, i.e., it does not vanish beyond any
specific border, and may accumulate amongst multiple concurrent
transmissions.


Only a few latest work have considered physical interference in the
context
\cite{jain2005iim,gronkvist2004tos,brar2006ces,behzad2007oil,chafekar2007cross,moscibroda2006complexity,moscibroda2006topology,chafekar2008approximation,moscibroda2007optimal},
etc. In \cite{jain2005iim}, Jain, et al. formulate the problem of
scheduling under physical interference model as an LP problem.
Unfortunately, no polynomial time solution and simulation-based
evaluation of scheduling is given. \cite{gronkvist2004tos} also
provides an exponential-time LP formulation. In \cite{brar2006ces},
Brar et al. present a heuristic scheduling method that is based on a
greedy assignment of weighted links. Although it is based on
physical interference model, the approximation factor of the
algorithm is given only when nodes are uniformly distributed. The
work of \cite{behzad2007oil} considers physical interference in the
minimum length scheduling problem. It uses a power-based
interference graph model, which describes the interference
relationship of every two links according to the SINR of the
receiver. However, the model fails to consider the accumulation
effect of interference. In \cite{chafekar2007cross}, approximation
algorithms for packet scheduling to minimize end-to-end delay with
physical interference model are proposed. The works of
\cite{moscibroda2006complexity,moscibroda2006topology} study the
problem of scheduling edges with SINR constraints to ensure that
some property (e.g., connectivity) is satisfied. Similar with the
work of \cite{chafekar2008approximation,moscibroda2007optimal}, they
take power control, scheduling or routing together into account.

The work of \cite{friderikos2006non} is similar to our centralized
algorithm discussed in section \ref{section4}. It also formulates
the scheduling problem as ILP, but the randomized rounding procedure
uses an iteration method which is more time-consuming than ours.
Because the rounding procedure would not stop until a feasible
solution is found, it costs a very long time to converge, and in
some cases, it may not converge at all.

Various distributed algorithms have been proposed for finding good
approximations of the scheduling problem based on the protocol
interference model (e.g.
\cite{chaporkar2005throughput,lin2005impact,modiano2006maximizing },
etc.). Only a few previous work propose distributed algorithm based
on the physical interference model. The work of
\cite{yi2007optimal,yi-mac,yi2006learning} propose distributed
algorithms based on the physical interference model, which is
lattice-throughput-optimal. But the approximation ratio is not
given. The work of \cite{joo2007performance} develops a
constant-time distributed random access algorithm for scheduling and
gives the performance bound of the algorithm. In
\cite{scheideler2008log}, a distributed and randomized protocol is
proposed, which uses physical carrier sensing to reduce message
overhead. It is similar to our approach discussed in section
\ref{section5}. But the protocol in \cite{scheideler2008log} is
sensitive to the scale of the network and spends much time to learn
this information.

\section{Network and interference models}\label{section3}
We abstract a wireless multi-hop network as a directed graph
$G(V,E)$ where $V$ is a set of vertices denoting the nodes
comprising the network and $E$ is a set of directed edges between
vertices representing inter-node wireless links. The Euclidean
distance between any two nodes $v_{i},v_{j}\in V$, is denoted by
$d(v_{i},v_{j})$. Let $e_{ij}\in E $ denotes the edge between $v_i$
and $v_j$.

Each node is assumed to be equipped with a single transceiver
working in the half-duplex way, and all nodes share a common
channel. So a node can not send and receive packets simultaneously.
All antennas  are omnidirectional.

It is assumed that the network is using Time Division Multiple
Access (TDMA) MAC protocol. The time is divided into slots of fixed
length, which are grouped into frames. To increase capacity, spatial
reuse TDMA (STDMA) \cite{nelson1985stc} can be used, which is an
extension of TDMA. The capacity is increased by spatial reuse of the
time slots.

We use the physical interference model \cite{gupta2000cwn} to
describe the interferences between active links. In this model, the
successful reception of a transmission depends on the received
signal strength, the interference caused by nodes transmitting
simultaneously, and the ambient noise level. The received power
$P_{r}(s_{i})$ of a signal transmitted by sender $s_{i}$ at an
intended receiver $r_{i}$ is
\begin{displaymath} \label{power of
receiver}
    P_{r}(s_{i}) = P(s_{i})\cdot g(s_{i},r_{i}),
\end{displaymath}
where $P(s_{i})$ is the transmission power of $s_{i}$ and
$g(s_{i},r_{i})$ is the propagation attenuation (link gain) modeled
as $g(s_{i},r_{i}) = d(s_{i},r_{i})^{-\alpha}$. The path-loss
exponent $\alpha$ is a constant between 2 and 6, whose exact value
depends on external conditions of the medium (humidity, obstacles,
etc.), as well as the exact sender-receiver distance. As common, we
assume that $\alpha > 2$ \cite{gupta2000cwn}. Given a
sender-receiver pair $(s_i, r_i)$, we use the notation $I_r(s_j) =
P_r(s_j)$ for interference from any other sender $s_j$ concurrent to
$s_i$. The total interference $I_r$ at the receiver $r_i$ is the sum
of the interference power caused by all nodes that transmit
simultaneously, except the intending sender $s_i$. Mathematically,
we have $I_r := \sum_{s_j\in V\backslash\{ s_i \}} I_r(s_j)$.
Finally, let $N$ denote the ambient noise power level. Then, $r_i$
receives transmission successfully from $s_i$ if and only if
\begin{eqnarray} \label{SINR}
    SINR(r_i)&= &\frac{P_r(s_i)}{N+\sum_{s_j\in
V\backslash\{ s_i \}} I_r(s_j)} \nonumber \\
             &= &\frac{P(s_{i})\cdot g(s_{i},r_{i})}{N+\sum_{s_j\in V\backslash\{
s_i \}} P(s_{j})\cdot g(s_{j},r_{i})} \nonumber \\
             &= &\frac{\frac{P(s_i)}{d(s_{i},r_{i})^{-\alpha}}}{N+\sum_{s_j\in V\backslash\{
s_i \}} \frac{P(s_{j})}{d(s_{j},r_{i})^{-\alpha}}} \geq \beta
\nonumber,
\end{eqnarray}
where $\beta$ is the minimum $SINR$ threshold required for a
successful message reception.

\section{The centralized approximation algorithm}\label{section4}

In this section, we first formulate the link scheduling problem
under the physical interference model as an Integer Linear
Programming problem, then give an approximate algorithm using
randomized rounding method, which can be done in polynomial time.
\subsection{Problem formulation}
We denote by $V_T\subseteq V$ and $V_R\subseteq V$ the set of
transmitting and receiving nodes respectively. The time is divided
into slots of fixed length and a frame is composed of $T$ times
slots, the length of which is constant. In time slot $t\in [1,T]$,
there exists at least one edge being scheduled to transmit. We
denote by $b_{ij} (e_{i,j}\in E)$ the total traffic rate through
link $(i,j)$ and we use these to measure throughput. The goal of the
scheduling method is to transmit all the edges in a frame to gain
the maximum throughput.

To formally formulate the problem, the boolean variables $x_{ij}^t$
is defined as
\begin{displaymath} \label{blean varible x}
x_{ij}^t = \left\{
\begin{array}{ll}
1 & \mbox{if link}\,\, e_{ij} \,\,\mbox{is scheduled to be transmit in slot}\,\,$t$\\
0 & \mbox{otherwise}
\end{array}
\right.
\end{displaymath}

Based on above assumptions, the linear integer programming
formulation can be written as follows,
\begin{displaymath} \label{object function}
    max\,\, \frac{1}{T}\sum_{t=1}^{T}\sum_{e_{ij}\in E} b_{ij}x_{ij}^t
\end{displaymath}
\begin{equation} \label{constraint_1}
    s.t.\,\,\,\,\,\,\,\,\,\,\,\,\sum_{t=1}^{T} x_{ij}^t \geq 1, \,\,\forall e_{ij}\in E
\end{equation}
\begin{equation} \label{constraint_2}
    \sum_{v_i\in V^T} x_{ij}^t \leq 1, \,\,\forall v_j\in V^R, \forall t
\end{equation}
\begin{equation} \label{constraint_3}
    \sum_{v_j\in V^R} x_{ij}^t \leq 1, \,\,\forall v_i\in V^T, \forall t
\end{equation}
\begin{equation} \label{constraint_4}
    \sum_{v_j\in V^T} x_{ij}^t +\sum_{v_k\in V^T} x_{ki}^t \leq 1, \,\,\forall v_i\in V^S\cap
    V^R, \forall t
\end{equation}
\begin{equation} \label{constraint_5}
    \frac{\frac{P(v_i)}{d(v_{i},v_{j})^{-\alpha}}x_{ij}^t+(1-x_{ij}^t)\Delta}{N+\sum_{e_{kj}\in E\backslash \{e_{ij}\}} {\frac{P(v_{k})}{d(v_{k},v_{j})^{-\alpha}}x_{kj}^t}} \geq \beta,
\,\,\forall e_{ij}\in E, \forall t
\end{equation}
\begin{equation} \label{constraint_7}
    x_{ij}^t \in \{0,1\}, \,\,\forall e_{ij}\in E, \forall t
\end{equation}
The objective function of the formulation is to maximize the total
network throughput, which is defined as the total traffic
transmitted per slot. Constraint (\ref{constraint_1}) guarantees
that each active edge should be scheduled at least once. Constraint
(\ref{constraint_2}) and constraint (\ref{constraint_3}) make sure
that each node can only receive or send signal from or to one
another node. This is because each node has only one transceiver.
Under constraint (\ref{constraint_4}), each node can not send and
receive at the same time because it work in a half-duplex way.
Constraint (\ref{constraint_5}) expresses the required SINR
threshold that should be satisfied in order to have a successful
reception at the receiver. The item $(1-x_{ij}^t)\Delta$ ensures
that the inequality is also satisfied when link $e_{ij}$ is not
scheduled in time slot $t$ (i.e. $x_{ij}^t=0$), for a sufficiently
large value of $\Delta$.
\subsection{Approximate algorithm with randomized rounding}

Since ILP problems are NP-complete, there is no efficient algorithm
known for solving them in bounded time (and there can not exist any
unless $P=NP$). The above formulation does not help us solve the
scheduling problem. However, it does guide us to a natural
relaxation which helps to find a good approximate algorithm.
\subsubsection{Randomized rounding procedure}

We can use randomized rounding method to find efficient and near
optimal solutions \cite{raghavan1987rrt}. In this procedure, each
variable $x_{ij}^t$ will be relaxed to be in the range from zero to
one and therefore converted from integer variables to fractional
ones, i.e., $x_{ij}^t \in [0,1], \,\, for \,\,\forall e_{ij}\in E,
\forall t$. The ILP problem is then relaxed to be a linear
programming problem, which is called LP-relaxation. The optimal
solution of the LP-relaxation can be treated as a probability vector
with which we choose to include a link to a specific times lot.
Denote the optimal solution as $\hat{x}_{ij}^t$. In that sense, if
$\hat{x}_{ij}^t$ is near one, it is likely that this link will be
included in the current times lot. If, on the other hand, the
optimal solution of the LP-relaxation problem is near zero, it will
probably not be included in the current times lot. More
specifically, the rounding procedure can be described as follows,
\begin{equation} \label{blean varible x}
x_{ij}^t = \left\{
\begin{array}{ll}
1, & \mbox{with probability}\,\, \hat{x}_{ij}^t\\
0, & \mbox{with probability}\,\, 1-\hat{x}_{ij}^t
\end{array}
\right.
\end{equation}

However, after the rounding procedure, all the constraints may not
be satisfied. For example, we suppose in constraint
(\ref{constraint_2}), for $\forall v_j\in V^R, \forall t$, there are
two nodes $v_{1}, v_{2}\in V_{T}$. If $\hat{x}_{1j}^t = 0.3$ and
$\hat{x}_{2j}^t = 0.7$, constraint (\ref{constraint_2}) is
satisfied. But after the rounding procedure, $x_{1j}=1$ with
probability 0.3, and $x_{2j}=1$ with probability 0.7. So
$\sum_{v_i\in V^T} x_{ij}^t = 2$ with probability 0.21. That is with
probability 0.21, constraint (\ref{constraint_2}) is not satisfied.
To overcome this problem, the work in \cite{friderikos2006non} uses
a iteration method. The rounding procedure would not stop until a
feasible solution is found, which costs a very long time to
converge. Even worse, it may not find any feasible solution in the
end. Unlike the work in \cite{friderikos2006non}, we adjust the
results of the rounding procedure based on all the constraints, and
no iteration procedure is needed. The detail of our proposed
algorithm is discussed in the following part.
\subsubsection{Approximate algorithm}
we denote by $\hat{\phi}^t=\{e_{ij}:x_{ij}^t=1\}$ the set of edges
that will be transmitted after the rounding procedure at time slot
$t$. As discussed above, all the constraints of the ILP program may
not be satisfied. So the solution of the rounding procedure will be
adjusted in algorithm 1 to satisfy all the constraints. As the
solution of the relaxed LP problem $\{\hat{x}_{ij}^t:\forall
e_{ij}\in E,t\in T\}$ is the probability for the wireless links to
transmit, we reorder the elements in $\hat{\phi}^t$ according to the
probabilities in the non-decreasing order.
\begin{algorithm}[!tb]
\caption{Centralized Approximate Algorithm} \label{Centralized
algorithm}
\begin{algorithmic}[1]
\STATE Consider all links in $E$:
\STATE Relax the integer linear programming problem to be a linear programming problem
(LP-relaxation);
\STATE Find the solutions of the relaxed LP
$\{\hat{x}_{ij}^t:\forall e_{ij}\in E,t\in T\}$;
\STATE Do randomized rounding and get the rounding solution
$\{x_{ij}^t:
\forall e_{ij}\in E,t\in T\}$ from equation (\ref{blean
varible x});
\FOR{each $t\in T$} \STATE Get the set
$\hat{\phi}^t=\{e_{ij}:x_{ij}^t=1\}$;
\STATE Consider all links $e_{ij}\in \hat{\phi}^t$ in non-decreasing order of
$\hat{x}_{ij}^t$:
\IF{constraint \ref{constraint_2} or constraint
\ref{constraint_3} or constraint \ref{constraint_4} is not
satisfied}
\STATE $x_{ij}^t=0$;
\STATE Move $e_{ij}$ away from
$\hat{\phi}^t$;
\ENDIF
\STATE Consider all links $e_{ij}\in
\hat{\phi}^t$ in non-decreasing order of $\hat{x}_{ij}^t$:
\IF{constraint \ref{constraint_5} is not satisfied}
\STATE $x_{ij}^t=0$;
\STATE Move $e_{ij}$ away from
$\hat{\phi}^t$;
\ENDIF
\ENDFOR

\STATE Consider all links $e_{ij}\in E$:
\FOR{each link $e_{ij}\in E$} \IF{$\sum_{t=1}^{T} x_{ij}^t < 1$}
\STATE Consider the biggest element $\hat{x}_{ij}^t$ for all $t\in T$:
\STATE $\hat{x}_{ij}^t=1$;
\STATE Execute from line (5) to (17)
\ENDIF
\ENDFOR

\end{algorithmic}
\end{algorithm}

Algorithm \ref{Centralized algorithm} proceeds in two phases: phase
1 is corresponding to line 1 to 4; while phase 2 is corresponding to
the rest of the algorithm. In phase 1, the ILP problem is relaxed to
be a linear programming problem which is solved by the randomized
rounding procedure. As mentioned before, there exists a problem that
the rounding solution $\{x_{ij}^t:\forall e_{ij}\in E,t\in T\}$ may
not satisfy all the constraints in the ILP problem. It can be
resolved in phase 2. From line 5 to 17, it is checked slot by slot
whether constraints (2) to (5) are satisfied. The solution of the
relaxed LP $\{\hat{x}_{ij}^t:\forall e_{ij}\in E,t\in T\}$ can be
viewed as the probability of the transmission of the links. The
links with larger value of $\hat{x}_{ij}^t$ have higher priority to
transmit. So we check the links in a non-decreasing order of
$\hat{x}_{ij}^t$. In each slot, if a link doesn't satisfy any of the
constraints of (2),(3) or (4), it will not transmit in the current
slot. After all links in $\hat{\phi}^t$ are checked, constraint (5)
is checked from line 12 to 16. Because constraint (5) is the
interference constraint, with which the transmission of a link is
related to all the other transmitting links, it is checked after
constraint (2) to (4) being checked. Constraint (1) guarantees that
each active link should be scheduled at least once in a frame.
However, after the procedure of randomized rounding and the check of
constraint (2) to (4), $x_{ij}^t$ of a link $e_{ij}$ may be rounded
to 0, or adjusted from 1 to 0, leading that constraint (1) is not
satisfied. From line 18 to 25, constraint (1) is check. For a link
$e_{ij}$ that doesn't satisfy constraint (1), find the time slot $t$
in which it has the largest probability to transmit (line 21). Then
let $e_{ij}$ has probability 1 to transmit, which guarantees that
$e_{ij}$ is scheduled once. Constraint (1) is then satisfied.
\subsubsection{Performance analysis}
In this part, the performance of algorithm \ref{Centralized
approximation algorithm} is analyzed, including the complexity and
the approximation ratio. It can be seen that algorithm
\ref{Centralized algorithm} can finish in polynomial time, and the
probability lower bound of being a $(1-\theta)$-approximation
algorithm ($\forall \,0<\theta < 1$) is calculated.

The complexity of phase 1 depends on the algorithm solving the
LP-relaxed problem. Let it be denoted by $O(P)$. Since it is a
linear programming problem, it runs in polynomial time. Thus $P$ is
polynomial. Let $N$ be the total number of the links in $E$. It can
be easily seen that the complexity from line 5 to 17 in algorithm 1
is $O(NT)$, because there are two loops there. Similarly, the
complexity from line 18 to 25 is $O(N^2T)$. Thus the complexity of
the whole algorithm is $O(max(P,N^2T))$, which is polynomial.

The approximate ratio of algorithm \ref{Centralized approximation
algorithm} can be got from the following theorem:

\newtheorem{theorm}{Theorem}
\begin{theorm}[Approximate ratio of the centralized algorithm]\label{theorem 1}
For $\forall 0< \theta <1$ and $-\theta<\Delta A/\hat{A}<1-\theta$,
the probability of algorithm \ref{Centralized approximation
algorithm} being $(1-\theta)$-approximate to the optimization is
lower bounded by $1-e^{-\frac{(\theta + \frac{\Delta
A}{\hat{A}})^2\hat{A}}{2}}$, where $\hat{A}$ is throughput
calculated by the LP-relaxation, $\Delta A$ is the variation of the
throughput in phase 2.
\end{theorm}
\begin{IEEEproof}
In phase 1, let $A_{rand}$ and $\hat{A}$ be total throughput
calculated by the randomized rounding in phase 1 and LP-relaxation
respectively, i.e.
$A_{rand}=\frac{1}{T}\sum_{t=1}^{T}\sum_{e_{ij}\in E}
b_{ij}x_{ij}^t$, and
$\hat{A}=\frac{1}{T}\sum_{t=1}^{T}\sum_{e_{ij}\in E}
b_{ij}\hat{x}_{ij}^t$. From equation (\ref{blean varible x}), we can
get $E(x_{ij}^t)=\hat{x_{ij}^t}$. Thus,
\begin{eqnarray}
E(A_{rand})&=&E(\frac{1}{T}\sum_{t=1}^{T}\sum_{e_{ij}\in E} b_{ij}x_{ij}^t)\nonumber\\
           &=&\frac{1}{T}\sum_{t=1}^{T}\sum_{e_{ij}\in E}E(b_{ij}x_{ij}^t) \nonumber\\
           &=&\hat{A}\label{E(A_rand)}
\end{eqnarray}
According to equation (\ref{E(A_rand)}) and Chernoff Bound
\cite{raghavan1987rrt}, we can get the following bound,
\begin{eqnarray}
Pr(A_{rand}\geq(1-\delta)\hat{A}) &=1-Pr(A_{rand}<(1-\delta)\hat{A}) \nonumber\\
                                  &\geq 1-e^{-\frac{\delta^2\hat{A}}{2}},\,\,
                                  0<\delta<1 \label{chernoff bound}
\end{eqnarray}
Let $A$ be the final solution of algorithm \ref{Centralized
approximation algorithm} and $A_{Opt}$  be the optimal solution of
the ILP problem. Obviously,
\begin{equation}\label{A_OPT}
A\leq A_{Opt} \leq \hat{A}
\end{equation}
In phase 2, the randomized rounding solution is checked whether it
satisfies all the constraints in the ILP problem, and some
adjustments are made, leading that the throughput changes. Suppose
the throughput changes by $\Delta A$ after phase 2, and it follows
that
\begin{equation}\label{A=A+delta A}
A=A_{rand}+\Delta A
\end{equation}
Substituting equation (\ref{A=A+delta A}) into (\ref{chernoff
bound}), we can derive
\begin{equation}\label{bound_1}
Pr(A\geq(1-\delta+\frac{\Delta A}{\hat{A}})\hat{A})\geq
1-e^{-\frac{\delta^2\hat{A}}{2}},\,\, 0<\delta<1
\end{equation}
In phase 2, the worst case is that all links are adjusted not to
transmit, leading the throughput $A=0$, i.e. $A_{rand}+\Delta A =
0$; while the best case is that after the adjustment, the solution
approaches to the optimal, leading $A=A_{rand}+\Delta A\leq
\hat{A}$. Thus we can get the variation range of $\Delta A$ as
follows,
\begin{equation}\label{Delta A}
-A_{rand}\leq \Delta A \leq \hat{A}-A_{rand}
\end{equation}
Let $\theta=\delta-\frac{\Delta A}{\hat{A}}$. Since $0<\delta<1$, it
can be derived that $-\theta<\Delta A/\hat{A}<1-\theta$. Equation
(\ref{bound_1}) can be written as follows,
\begin{equation}\label{bound_2}
Pr(A\geq(1-\theta)\hat{A})\geq 1-e^{-\frac{(\theta + \frac{\Delta
A}{\hat{A}})^2\hat{A}}{2}},
\end{equation}
where $-1+\frac{A_{rand}}{\hat{A}}\leq\theta \leq
1+\frac{A_{rand}}{\hat{A}}$ (from equation (\ref{Delta A})), and
$-\theta<\Delta A/\hat{A}<1-\theta$.

Because $(0,1)\subset
[-1+\frac{A_{rand}}{\hat{A}},1+\frac{A_{rand}}{\hat{A}}]$, according
to equation (\ref{A_OPT})(\ref{bound_2}), we can get
\begin{eqnarray}
Pr(A\geq(1-\theta)A_{Opt})&\geq Pr(A\geq(1-\theta)\hat{A})\nonumber \\
                          &\geq 1-e^{-\frac{(\theta + \frac{\Delta A}{\hat{A}})^2\hat{A}}{2}}
\end{eqnarray}
where $\theta \in (0,1)$, and $-\theta<\Delta A/\hat{A}<1-\theta$.

\end{IEEEproof}

\section{The distributed algorithm}\label{section5}
The approximate algorithm discussed in section \ref{section4} uses
the physical interference model to represent the interference and
has a good performance. However, as mentioned in section
\ref{section1}, the centralized algorithm requires the global
information of the network to do the scheduling. Collecting the
required information may causes a lot of bandwidth overhead,
especially when the network scale is large. Moreover, in some cases,
the information is hard to get. The difficulties in the
implementation of centralized algorithm call for the distributed
scheduling scheme that may achieve a fraction of the overall optimal
performance.

In this part,  we extend the centralized algorithm \ref{Centralized
approximation algorithm} to a distributed solution, which takes the
physical interference model into account and implements in three
phases. It is assumed that nodes in the network can perform physical
carrier sensing, and they can set the carrier sensing range to
different values. As we show in the following part, by properly
tuning the carrier sensing range, all the constraints of the ILP
problem in section \ref{section4} can be satisfied. The
approximation ratio is also proved.

\subsection{Carrier sensing range calculation}
Let $d_{max}$ and $d_{min}$ be the maximum and minimum link length
in the network respectively. The length diversity $k$ is defined as
\begin{equation}\label{diversity}
k=\lfloor \log{(d_{max}/d_{min})} \rfloor
\end{equation}
Suppose each node transmits at the same power $P$. As the ambient
noise power is much less than the interference power, it is omitted
here. The following theorem describes how to tune the carrier
sensing range for each node.
\begin{figure}[!h]
\centering
\includegraphics[width=2.5in]{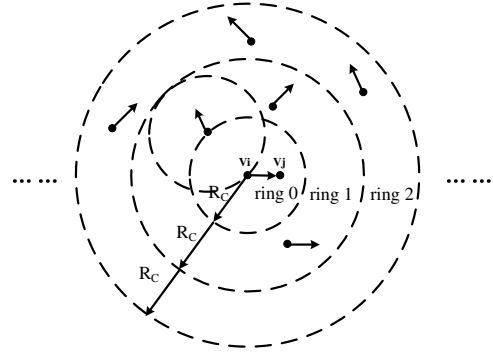}
\caption{The Euclidean plane is divided into a series of rings}
\label{Fig:the Euclidean plane is divided into a series of rings}
\end{figure}

\begin{theorm}[Carrier sensing range]\label{theorem 2}
If the carrier sensing range is set to be $R_C=\rho 2^k$, the
transmitting links in the network will not be interfered by all the
other links, where $\rho$ is defined as
\begin{equation}\label{definition of rou}
\rho=4(2\pi \beta \frac{\alpha -1}{\alpha-2})^\frac{1}{\alpha},
\end{equation}
$\alpha$ is the path-loss exponent and $\beta$ is the minimum $SINR$
threshold).
\end{theorm}
\begin{IEEEproof}
Without loss of generality, suppose link $e_{ij}\in E$ is
transmitting data from node $v_{i}$ to $v_{j}$. we normalize the
minimum link length $d_{min}$ to be 1. From the definition of the
length diversity, we can get $2^k\leq d_{max} \leq 2^{k+1}$. The
length of link $e_{ij}$ satisfies $d_{ij}\leq{d_{max}
\leq{2^{k+1}}}$, thus the perceived power at $v_j$ from $v_i$ is at
least
\begin{equation}\label{received power}
P_{ij}\geq{\frac{P}{2^{(k+1)\alpha}}}
\end{equation}
The plane can be divided into a series of rings with the center
located at sending node $v_i$, which can be seen in figure
\ref{Fig:the Euclidean plane is divided into a series of rings}.
They are denoted as ring $0$, $1$, $\cdot\cdot\cdot$, respectively.
Since the region of ring $0$ is in the carrier sensing range of the
sending node $v_i$, there are no other sending nodes locating in
ring $0$. For a sending nodes $v_m$ located in ring $1$, it must be
at least $d_{mj}\geq R_C-2^{k+1}$ away from $v_j$. Since each
sending node must be outside the carrier sensing range of other
sending nodes, there are at most $\pi /\arcsin(\frac{1}{2})=6$
sending nodes can transmit concurrently in ring 1. Consequently, the
aggregated interference at $v_j$ is
\begin{eqnarray}\label{interference in annular area 1}
\sum_{l=1}^{6} P_{lj} &=&\sum_{l=1}^{6} \frac{P}{d_{lj}^\alpha}\nonumber \\
                      &\leq &\frac{6P}{(R_C-2^{k+1})^\alpha}\nonumber
\end{eqnarray}
In ring $2$, the sending nodes are at least $2R_C-2^{k+1}$ away from
$v_j$. there are at most $\pi /\arcsin(\frac{1}{4})\leq
\frac{\pi}{1/4}=4\pi$ sending nodes can transmit concurrently. So
the aggregated interference at $v_j$ is
\begin{displaymath} \label{interference in annular area 2}
   \sum_{l=1}^{4\pi} P_{lj} \leq{\frac{4\pi P}{(2R_C-2^{k+1})^\alpha}}
\end{displaymath}
Similarly, in ring $m$, the sending nodes are at least
$mR_C-2^{k+1}$ away from $v_j$. there are at most $\pi
/\arcsin(\frac{1}{2m})\leq \frac{\pi}{1/2m}=2m\pi$ sending nodes can
transmit concurrently. So the aggregated interference at $v_j$ is
\begin{displaymath} \label{interference in annular area m}
   \sum_{l=1}^{2m \pi} P_{lj} \leq{\frac{2m\pi P}{(mR_C-2^{k+1})^\alpha}}
\end{displaymath}
Consequently, the accumulated interference at the receiving node
$v_j$ is upper bounded by
\begin{eqnarray}\label{total interference}
I_{v_j} &\leq &\sum_{m=1}^{\infty} \frac{2m\pi P}{(mR_C-2^{k+1})^\alpha}\nonumber \\
        &=&\sum_{m=1}^{\infty} \frac{2m\pi P}{(m\rho 2^k-2^{k+1})^\alpha}\nonumber \\
        &=&\frac{2\pi P}{2^{k\alpha}}\sum_{m=1}^{\infty} {\frac{m}{(m\rho -2)^\alpha}} \label{x-2>x/2}\\
        &\leq &\frac{2\pi P}{2^{k\alpha}}\sum_{m=1}^{\infty} {\frac{m}{(m\rho /2)^\alpha}} \nonumber\\
        &=&\frac{2\pi P}{2^{k\alpha}(\rho /2)^\alpha}\sum_{m=1}^{\infty} {\frac{1}{m^{\alpha-1}}}\nonumber \\
        &\leq &\frac{2\pi
        P}{2^{(k-1)\alpha}\rho^\alpha}\frac{\alpha-1}{\alpha-2}\label{Riemann'zeta function}
\end{eqnarray}
where (\ref{x-2>x/2}) follows because $x-2>x/2$, $\forall x>4$ and
$\rho
>4$, given that $\beta \geq 1$ and $\alpha \geq 2$; and (\ref{Riemann'zeta function}) follows
from a bound on Riemann's zeta function
\cite{goussevskaia2007complexity}. From equation (\ref{definition of
rou}), (\ref{received power}) and (\ref{Riemann'zeta function}), we
can get the $SINR$ at the receiving node $v_j$ as follows:
\begin{eqnarray}\label{SINR}
SINR_{v_j} &=&\frac{P_{ij}}{I_{v_j}}\nonumber \\
           &\geq &\frac{\frac{P}{2^{(k+1)\alpha}}}{\frac{2\pi P}{2^{(k-1)\alpha}\rho ^\alpha}\frac{\alpha-1}{\alpha-2}}\nonumber \\
           &=&\beta \nonumber
\end{eqnarray}

\end{IEEEproof}
\subsection{Algorithm description}
The frame structure is shown in Figure \ref{Fig:frame structure}. A
frame is composed of fixed length time slots. A time slot is divided
into 3 phases, namely, carrier sensing phase, RTS-CTS phase and
data-ack phase. Our distributed scheduling approach allows each node
to implement it in each of the three phases. Next, we will present
the details of the approach and show that after the three phases,
all the constraints in the ILP problem are satisfied.
\begin{figure}[!h]
\centering
\includegraphics[width=2.5in]{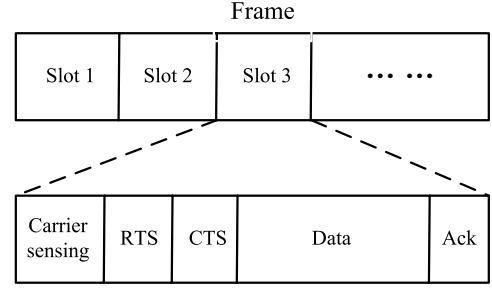}
\caption{Frame structure in distributed approach} \label{Fig:frame
structure}
\end{figure}
\subsubsection{Carrier sensing phase}
The purpose of this phase is to guarantee that there are no other
nodes sending data in the carrier sensing range of a transmitting
node. Each node in the network sets its carrier sensing range to be
$R_C$, which is calculated in theorem \ref{theorem 2}. So according
to theorem \ref{theorem 2}, the node will transmit its data without
any interference. Constraint (\ref{constraint_5}) in the ILP is
satisfied.

Consider a transmission slot starts at time t. The carrier sensing
phase begins at the beginning of each time slot. It lasts for a
period of $t_c$. Each node that wants to transmitting data randomly
selects a time $t_s$ and sends a \emph{SENSING} signal, which lasts
for a period of $\tau$ ($t\leq t_s \leq t+t_c-\tau$). The purpose of
the \emph{SENSING} signal is to let each node sense whether there
are other transmitting nodes in its carrier sensing range. If a node
do not sense any signal before sending the \emph{SENSING} signal, it
will occupy this time slot, and enter phase 2; otherwise, the node
randomly selects another time slot and waits to transmit again. So
the nodes located in each other's carrier sensing range will
transmit in different slots, and no interference will occur. For
each node, the purpose of randomly selecting a time to send
\emph{SENSING} signal is to prevent that all nodes send the
\emph{SENSING} signal concurrently, and they all will sense nothing.

If a link is scheduled, the sending node will not send the
\emph{SENSING} signal again until it can sense nothing, i.e., all
the sending nodes that have been scheduled at least once will wait
for the nodes that haven't been scheduled to transmit. This rule
guarantees that all the links are scheduled at least once in a
frame. So constraint (\ref{constraint_1}) is satisfied.
\subsubsection{RTS-CTS phase}
In this phase, the sending node  first sends RTS signal to the
corresponding receiving node. If the receiving node receives RTS, it
will respond CTS signal to the sending node. If the sending node
receives CTS signal, it will enter phase 3. Otherwise, it will
randomly select another time slot to transmit.

\begin{figure}[!h]
\centering
\includegraphics[width=2.5in]{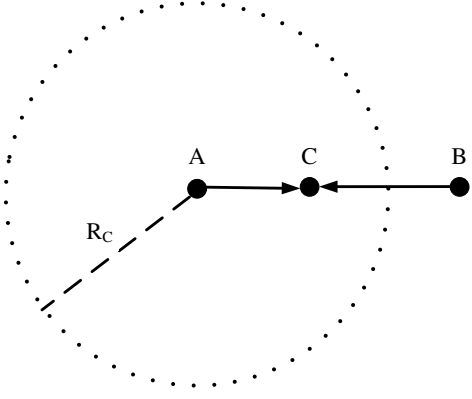}
\caption{An example in phase 2} \label{Fig:example in phase 2}
\end{figure}

This phase guarantees that constraint (\ref{constraint_2}) in the
ILP is satisfied, i.e., a node can not receive data from more than
one nodes simultaneously. An example can be seen in figure
\ref{Fig:example in phase 2}. Node A and B are both transmitting to
node C. Since node B is outside the carrier sensing range of node A,
A and B can transmit simultaneously in phase 1 without interference.
Because C can not receive data from more than one node
simultaneously (constraint \ref{constraint_2}), the data from A or B
will not be received. Through the RTS-CTS procedure, A or B will
transmit in another slot, and constraint (\ref{constraint_2}) is
satisfied.
\subsubsection{Data-ack phase}
In this phase, the sending node sends data to the receiving node,
and waits for the ack signal from the receiving node. If the
receiving node receives data correctly, it sends back ack signal to
confirm the correct reception. If the sending node doesn't receive
the ack signal, it will transmit the data again in the next slot.

Because of the distribution feature of our approach, a node can
decide not to send data to more than one receiving nodes
simultaneously. So constraint (\ref{constraint_3}) is satisfied.
Usually, the carrier sensing range is larger than the transmission
range. Thus, when a node is transmitting, no other nodes will
transmit to it. In addition, because of the half duplex character of
wireless nodes, a node will not send data to others when it is
receiving data. So constraint (\ref{constraint_4}) is satisfied. In
phase 1, constraint (\ref{constraint_1}) and (\ref{constraint_5})
are satisfied. Constraint (\ref{constraint_2}) is satisfied in phase
2. So all the constraints are satisfied in our distributed approach,
which can be viewed as a distributed solution of the ILP. Next, we
will analyze the approximation ratio of the distributed approach.
\subsection{Performance analysis}
The approximation ratio of the distributed approach can be described
in the following theorem:

\begin{theorm}[Approximation ratio of the distributed approach]\label{theorem 3}
The approximation ratio of the distributed approach is at most
$d_{max}^\alpha(\rho +2)^\alpha / \beta$, i.e., the distributed
approach is an $O(d_{max}^\alpha)$ approximate algorithm ($\rho$,
$\alpha$ and $\beta$ are the same as defined in theorem \ref{theorem
2}, and $d_{max}$ is the normalized maximum link length).
\end{theorm}
\begin{IEEEproof}
It can be seen that each sending node in the network can be located
in a certain circle with radius being the carrier sensing range
$R_C$. Without loss of generality, we consider a circular area
$\Phi_{v_i}$ with a sending node $v_i$ located in the center. Node
$v_i$ is sending data to $v_j$. Assume an optimal algorithm OPT can
schedule at most $q$ links in $T_{OPT}=1$ slot whose sending node is
located in $\Phi_{v_i}$. According to theorem \ref{theorem 2}, to
scheduling these $q$ links, our proposed approach needs $T\leq q$
time slots. So the approximation ratio is
\begin{displaymath}
\frac{T}{T_{OPT}} \leq q
\end{displaymath}

Next, we will calculate the maximum value of q. The best case is
that there are no node sending data outside $\Phi_{v_i}$, and all
the links in $\Phi_{v_i}$ can transmit successfully. For node $v_j$,
the perceived  SINR level is
\begin{eqnarray}
SINR(v_j)&=&\frac{P/d_{ij}^\alpha}{\sum_{k=1}^{q}
P/d_{kj}^\alpha}\nonumber \\
         &\leq &\frac{P}{q P/(\rho
         2^k+2^{k+1})^\alpha}\label{q}\\
         &=& SINR_{MAX}(v_j)\nonumber
\end{eqnarray}
where (\ref{q}) follows because $d_{ij}\geq 1$ and $d_{kj}\leq
2^{k+1}$. According to $SINR_{MAX}(v_j)\leq \beta$, we can get
\begin{eqnarray}
q &\leq & 2^{k\alpha}(\rho +2)^\alpha / \beta \nonumber \\
  &\leq & d_{max}^\alpha(\rho +2)^\alpha / \beta \label{d_max}\\
  &=& O(d_{max}^\alpha)\nonumber
\end{eqnarray}
where (\ref{d_max}) follows from the definition of diversity
(\ref{diversity}).
\end{IEEEproof}

From theorem \ref{theorem 3} we can see that the performance of the
distributed algorithm is closely related to the link length
diversity. The approach performs well when the diversity is small.
It can be seen distinctly from simulation (section \ref{section6}).
\section{Simulations}\label{section6}
In this section, we present the simulation results that verify the
performance of our proposed approximate algorithms in this paper.
\subsection{Simulation scenario}
We create a scenario where wireless nodes are uniformly distributed
in a $100\times 100$ square. In the simulation, we normalize the
transmission range of a node to be $R_T=1$, and the interference
range to be $R_I = 2.5$. For simplicity, all nodes are supposed to
transmit at the same power level. We also assume $N=-90dBm$, $\beta
=10dB$, and $\alpha=4$, which are similar with \cite{behzad2003pgb}
and \cite{yi2007optimal}. Let the frame length be $T=100$. The whole
simulation runs 100 times. In each simulation, $n$ communication
pairs are randomly selected. Suppose each of the selected links has
the same traffic rate which is normalized to be 1.
\begin{figure}[tbhp!]
\centerline{\subfigure[]{
\includegraphics[width=0.22\textwidth]{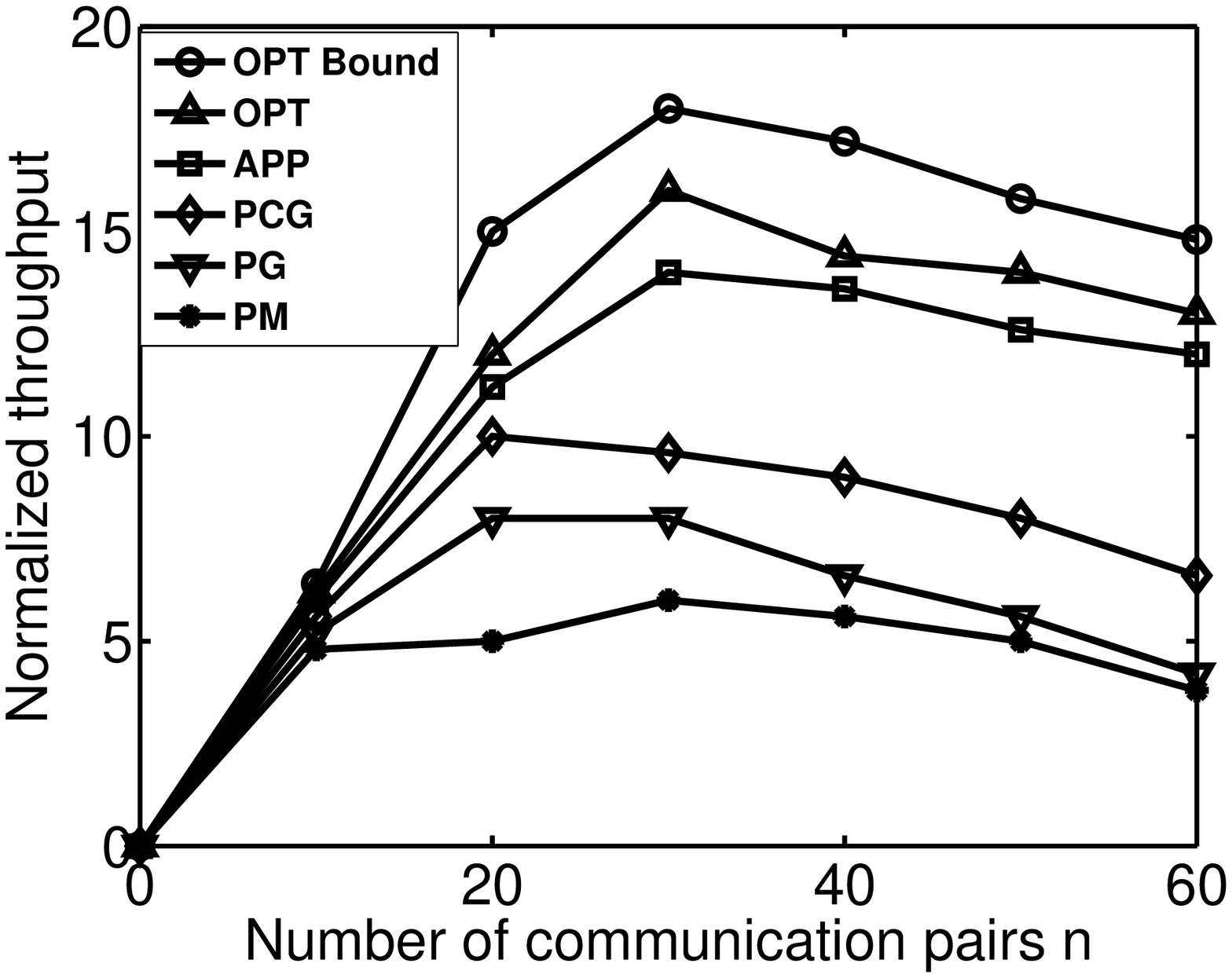}
\label{Subfig:comparation}} \hfil \subfigure[]{
\includegraphics[width=0.22\textwidth]{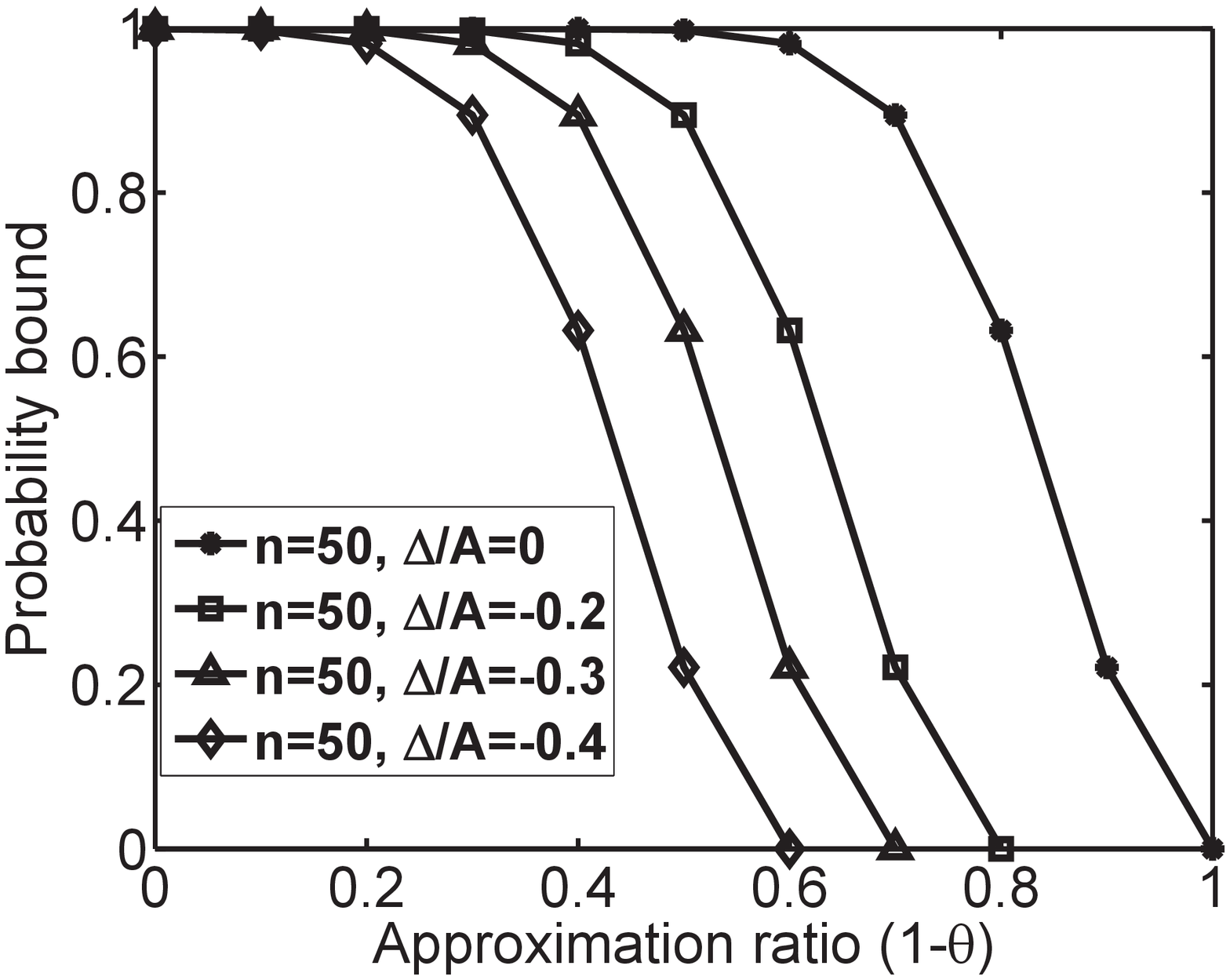}
\label{Subfig:probability bound versus delta/A}}}

\centerline{ \subfigure[]{
\includegraphics[width=0.22\textwidth]{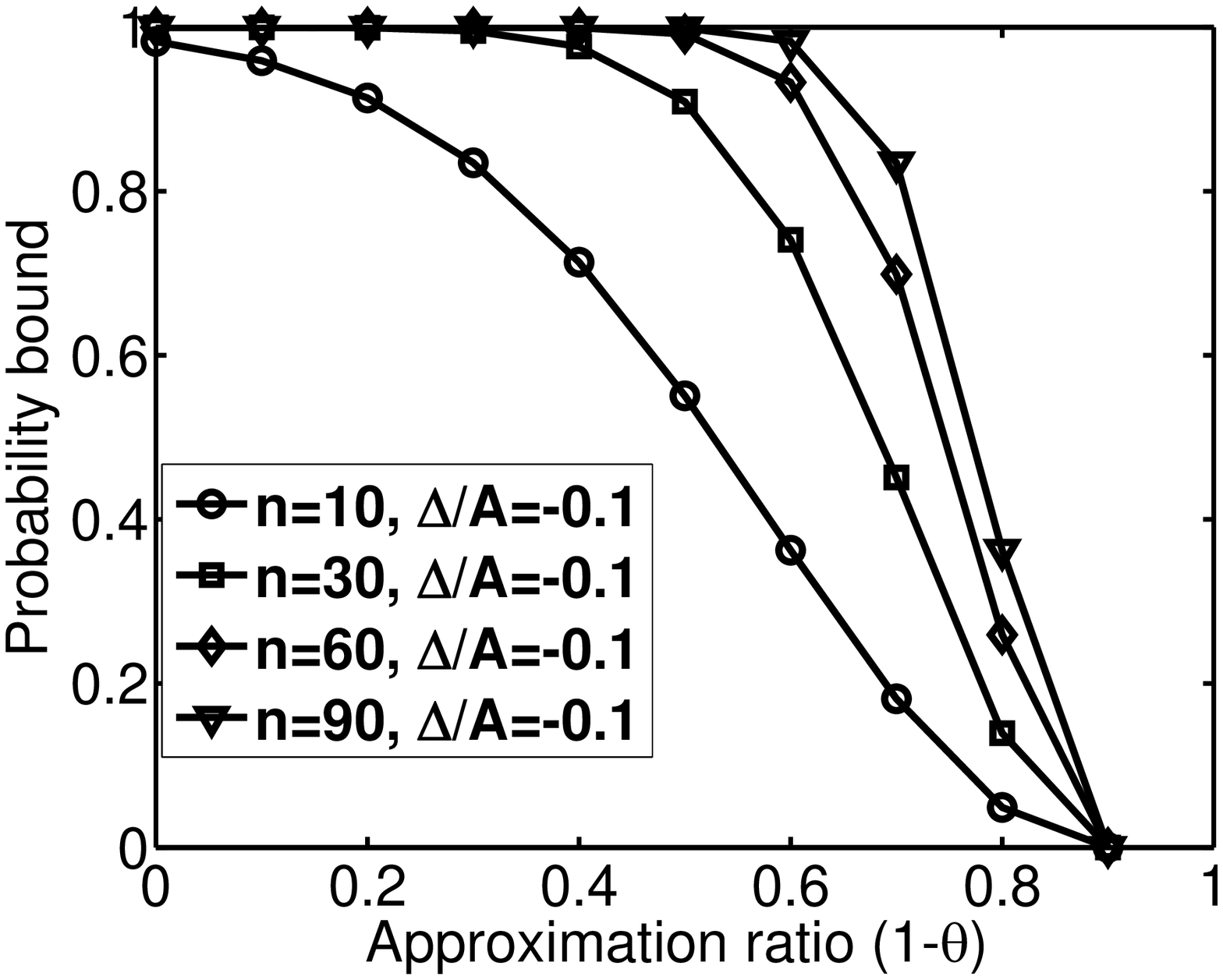}
\label{Subfig:probability bound versus n same delta/A}} \hfil
\subfigure[]{
\includegraphics[width=0.22\textwidth]{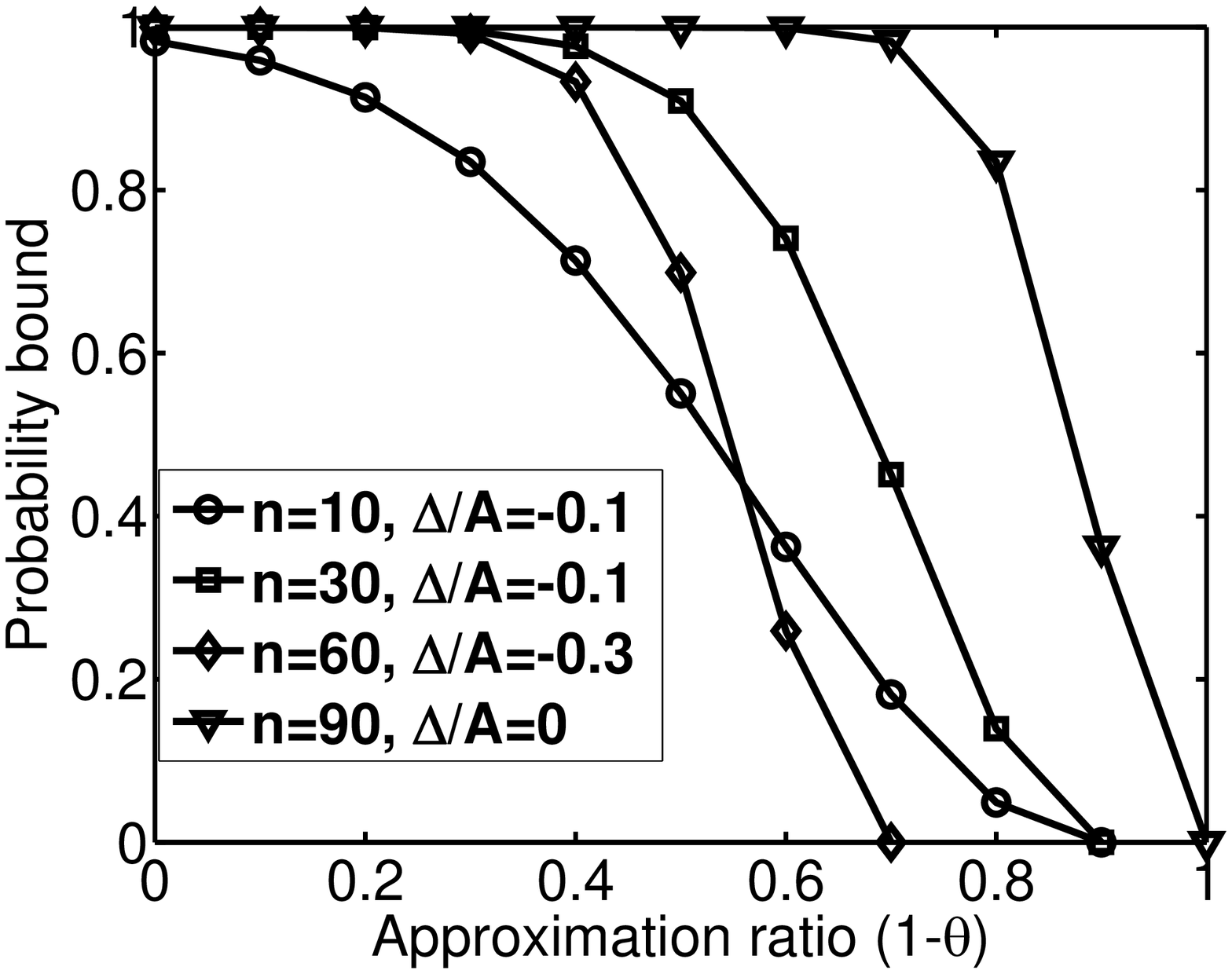}
\label{Subfig:probability bound versus n different delta/A}}}
\label{Fig:centrlized algorithm} \caption{The performance of the
centralized algorithm}
\end{figure}

%

\subsection{Centralized algorithm}
We first compare the throughput performance between our proposed
algorithm and several other centralized algorithms. The algorithm in
\cite{ramanathan1993sam} is based on protocol interference model and
a polynomial-time coloring algorithm is used. To express
conveniently, we call it Algorithm Protocol Model (PM). In
\cite{brar2006ces}, the algorithm is based on physical interference
model, which is the same as ours. It is a heuristic scheduling
method that is based on a greedy assignment of weighted links. We
call it Algorithm Physical Greedy (PG). The authors in
\cite{behzad2007oil} construct the conflict graph using the physical
interference model, which does not consider the accumulation effect
either. It is called Algorithm Physical Conflict Graph (PCG). The
throughput comparison of the several algorithms is shown in
Fig.\ref{Subfig:comparation}. The notation OPT Bound here stands for
the optimal solution of the LP-relaxation. The notation OPT is the
optimal solution of ILP, which is got by traversal searching.
Algorithm APP implies our approximate algorithm.

It can be seen that our proposed APP algorithm is well approximate
to the optimal solution of ILP, which is a little lower than the
upper bound. It also outperforms the protocol interference model
based algorithm (PM algorithm) and other physical interference model
based heuristic algorithms (PG and PCG algorithms). It performs ever
better when more nodes begin to transmit. The PM algorithm performs
the worst, because it uses the protocol interference model, which is
less precise than the physical interference model. When more and
more nodes begin to transmit, the throughput drops for all the
algorithms. This is because when interference is more severe, more
transmissions begin to conflict with each other.

Then we evaluate the influence of the parameter $\Delta A /\hat{A}$
(defined in theorem \ref{theorem 1}) to the ($1-\theta$)-approximate
lower bound. Here we denote $\Delta A /\hat{A}$ by $\Delta A /A$ in
the legends of Fig. 4. It can be seen from
Fig.\ref{Subfig:probability bound versus delta/A} that the lower
bound decreases while $\Delta A /\hat{A}$ decreases, which is from
the fact that more adjustments are made in the phase 2 of the
algorithm \ref{Centralized algorithm}. In our simulation, $\Delta A
/\hat{A}$ is rarely smaller than -0.3, implying that the
approximation ratio is larger than 0.5 with probability larger than
0.7, and the approximation ratio larger than 0.4 with probability
larger than 0.9.

When more links begin to transmit, more adjustments are made in
algorithm \ref{Centralized algorithm}, i.e., $\Delta A$ decrease.
The influence of the network scale $n$ to the performance can be
seen in Fig.\ref{Subfig:probability bound versus n same delta/A} and
\ref{Subfig:probability bound versus n different delta/A}. We can
get the following conclusions when $n$ increases:

$1)$ If $\Delta A /\hat{A}$ doesn't decrease, the probability lower
bound increases. In Fig.\ref{Subfig:probability bound versus n same
delta/A}, $\Delta A /\hat{A}$ is fixed to be 0.1. The probability
lower bound increases while more links begins to transmit. If
$\Delta A /\hat{A}$ increases, the probability lower bound also
increases (see plot ($n=60, \Delta A/A=-0.8$)and ($n=90, \Delta
A/A=0$) in Fig.\ref{Subfig:probability bound versus n same
delta/A}).

$2)$ If $\Delta A /\hat{A}$ decrease, the probability lower bound
dereases. It can be seen in Fig.\ref{Subfig:probability bound versus
n same delta/A}).




\subsection{Distributed algorithm}
In this simulation, nodes are not uniformly distributed, which is
different from the scenario in the centralized algorithm. We select
the transmission pairs properly to control the link length diversity
$k$.

\begin{figure}[!h]
\centering
\includegraphics[width=0.22\textwidth]{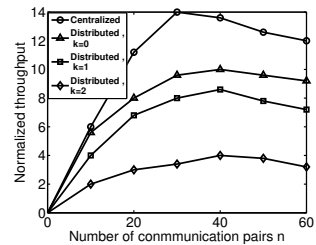}
\caption{The performance of distributed algorithm}
\label{Fig:Performance of distributed algorithm}
\end{figure}

In Fig. \ref{Fig:Performance of distributed algorithm}, the
throughput of the centralized algorithm and distributed approach are
compared. It can be seen that the distributed algorithm approaches
the centralized algorithm well when $k=0$. The throughput of $k=0$
and that of $k=1$ are very close to each other; while it decreases
sharply when $k=2$. It is because the approximation ratio is
$O(2^{k\alpha})$, which is exponential to $k$. However, in real
scenario $k$ is not very large, typically $k\leq 2$. The distributed
algorithm proposed in the paper is preferred to be implemented in
the scenario that the link length diversity is small. The ideal
scenario is that the nodes are located uniformly.
\section{Conclusion}\label{section7}
In this paper we study the link scheduling problem with the physical
interference model, which is more accurate than the protocol
interference model but notoriously hard to handle with. The
scheduling problem is formulated to be a ILP problem, and we propose
our centralized algorithm based on randomized rounding. We analyze
the performance of the algorithm and give the probability bound of
getting the guaranteed approximation ratio. As a distributed
solution of the ILP problem, we also propose a distributed approach,
which uses physical carrier sensing and implement in phases. The
approximation ratio is also given, which is associated with the link
length diversity.
%
%
%
%
\renewcommand{\baselinestretch}{1}
\selectfont
\section*{Acknowledgment}
This research is supported in part by the National Science
Foundation of China (Grant No.60672107, 60672142 and 60772053), the
Hi-tech Research and Development Program of China (Grant
No.2006AA10Z261, 2006AA10A301, and 2007AA100408), the China 973
Project (Grant No.2007CB307100 / 2007CB307105) and the China
Postdoctoral Science Foundation grants 20080430400.

\IEEEtriggeratref{15}


%

%
%


\end{document}